\documentclass[twocolumn,showpacs,preprintnumbers,amsmath,amssymb]{revtex4}

\usepackage{graphicx}
\usepackage{dcolumn}
\usepackage{bm}

\begin{document}

\title{Super-resolution and reconstruction of far-field ghost imaging via sparsity constraints}
\author{Wenlin Gong}
\email{gongwl@siom.ac.cn}
\author{Shensheng Han}
\affiliation{Key Laboratory for
Quantum Optics and Center for Cold Atom Physics of CAS, Shanghai
Institute of Optics and Fine Mechanics, Chinese Academy of Sciences,
Shanghai 201800, China}

\date{\today}

\begin{abstract}
For ghost imaging, the speckle's transverse size on the object plane
limits the system's imaging resolution and enhancing the resolution
beyond this limit is generally called super-resolution. By combining
the sparsity constraints of imaging target with ghost imaging
method, we demonstrated experimentally that super-resolution imaging
can be nonlocally achieved in the far field applying a new sparse
reconstruction method called compressive sensing. Some factors
influencing the quality of super-resolution ghost imaging via
sparsity constraints are also discussed.
\end{abstract}

\pacs{42.50.Ar, 42.30.Va, 42.30.Wb}
\maketitle Super-resolution is always an important topic in imaging
science \cite{Chaudhuri,Kolobov}. In practical applications, the
imaging resolution is limited by the noise and the bandwidth of the
system. Exploiting the evanescent components containing fine detail
of the electromagnetic field distribution at the object's immediate
proximity, super-resolution can be achieved, but this method is only
applied in the near-field range \cite{Ash,Ramakrishna,Pendry}. While
beyond the near-field range (namely in Fresnel and Fraunhofer
regions) \cite{Goodman}, the diffraction effect of the
transmitting/receiving system limits the imaging resolution, such as
scanning imaging, fluorescence imaging, telescope and so on
\cite{Kolobov,Goodman,Suarez}. Using additional a priori information
of optical system, the imaging resolution beyond Rayleigh
diffraction limit can be obtained. However, the improvement degree
is limited in practice because of the influence of detection noise
\cite{Kolobov,Goodman,Harris,Mallat,Hunt}. Ghost imaging (GI), which
is based on the quantum or classical correlation of fluctuating
light fields, has demonstrated theoretically and experimentally that
one can nonlocally image an object
\cite{Cheng,Gatti,Bennink,Gong,Zhang,Ferri,Valencia,Angelo,Gong1,Ferri1,Liu,Bache,Basano}.
Although differential ghost imaging \cite{Gong1,Ferri1} and the
spatial averaging technique \cite{Liu,Bache} can improve the
visibility of pseudo-thermal GI and speed up the convergence, the
imaging resolution is limited by the speckle's transverse size on
the object plane \cite{Gong,Zhang,Ferri}. When signals satisfied a
certain sparsity constraints, Donoho had demonstrated mathematically
that super-resolution restoration was possible \cite{Donoho} and
lots of sparse reconstruction methods had been used to reconstruct
the superresolved images \cite{Hunt,Park,DeGraaf,Donoho}. However,
the above sparse reconstruction methods were limited by their
special conditions. Recently, a new sparse reconstruction method
called compressive sensing (CS), which also relies on sparsity
constraints of images, has proved that images can be stably
extracted by random measurement when the sensing matrix satisfies
the restricted isometry property (RIP), and this method has been
widely applied in lots of fields to improve the signal-to-noise
ratio of images because it is robust to noise and universal, such as
data compression, magnetic resonance imaging, even ghost imaging
\cite{Katz,Candes1,Candes,Figueiredo}. For GI, the fluctuating light
field obeying Gaussian statistical distribution essentially
satisfies RIP and the measurement is also random. Therefore,
super-resolution GI is possible applying CS method because all
images are sparse in a proper representation basis \cite{Candes}.

\begin{figure}
\centerline{
\includegraphics[width=8.5cm]{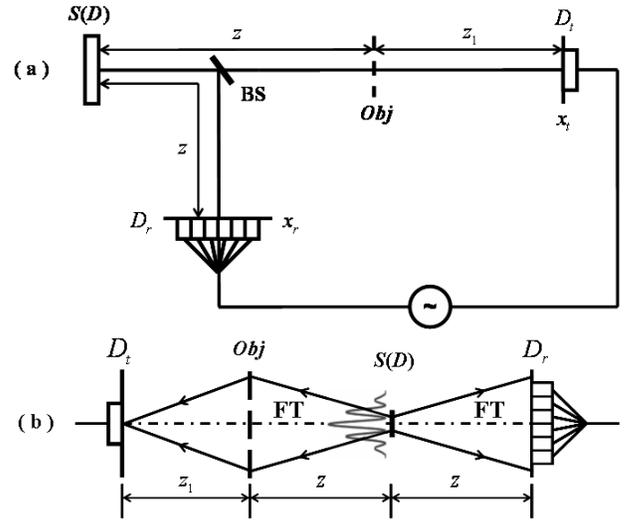}}
\caption{(a). Standard schematic of lensless far-field GI with
pseudo-thermal light; (b). the physical explanation of far-field GI,
the thermal source $S$ shown in the scheme (a) acts as a phase
conjugated mirror and a spatial low-pass filter because of its
finite transverse size.}
\end{figure}

For compressive ghost imaging demonstrated in Ref. \cite{Katz}, the
speckle's transverse size is small enough to resolve the object and
the test detector is positioned at the near field of the object,
thus the test detector should be a bucket detector, which can
collect the global information transmitted through the object. Fig.
1(a) presents the experimental schematic for lensless far-field
ghost imaging. Different from the case mentioned in Ref.
\cite{Katz}, the speckle's transverse size is too large to resolve
the object and the test detector is fixed in the far field of the
object, thus a single pointlike detector is enough to record the
global information from the object. In the experiment, the
pseudo-thermal source $S$, which is obtained by passing a focused
laser beam (with the wavelength $\lambda$=650nm and the source's
transverse size $D$) through a slowly rotating ground glass disk
\cite{Gong}, is divided by a beam splitter (BS) into a test and a
reference paths. In the test path, the light goes through a
double-slit (slit width $a$=100$\mu$m, slit height $h$=500$\mu$m and
center-to-center separation $d$=200$\mu$m) and then to a detector
$D_t$ fixed in the far field of the object (namely
$z_1>\frac{2d^2}{\lambda}$). In the reference path, the light
propagates directly to a camera $D_r$. Both the object and the
camera $D_r$ are located in the far field of the source (namely
$z>\frac{2D^2}{\lambda}$).

The intensity distribution on the detection plane at time $s$ can be expressed as \cite{Goodman}
\begin{eqnarray}
I_s (x, y)&=&\int {dx_1 } {dy_1 }{dx_2 } {dy_2 }E_s (x_1, y_1)E_s^*(x_2, y_2)\nonumber\\&\times& h^*(x, y; x_2, y_2)h(x, y; x_1, y_1).
\end{eqnarray}
where $E_s(x_1, y_1)$ denotes the light field on the source plane at time $s$, $h(x, y; x_1, y_1)$ and $h^*(x, y; x_2, y_2)$ are the impulse
function of optical system and its phase conjugate, respectively.

By Ref. \cite{Cheng,Gatti,Gong}, the correlation function for
far-field GI shown in Fig. 1(a) can be represented as
\begin{eqnarray}
G^{(2,2)} (x_r, y_r;x_t,y_t)\propto \left| \right.\int {dx' }  {dy' } T(x', y') \nonumber\\\times \sin c [\frac{D}{{\lambda z}}(x_r  - x')]
\sin c [\frac{D}{{\lambda z}}(y_r  - y')] \nonumber\\\times \exp \{ \frac{{j\pi }}{{\lambda z_1}}[(x_t  - x')^2+(y_t  -
y')^2]\}\nonumber\\\times \exp \{ \frac{{j\pi }}{{\lambda z}}({x'}^2-{x_r}^2+{y'}^2-{y_r}^2)\} \left. \right|^2 .
\end{eqnarray}
where $T(x, y)$ is the object's transmission function and $\sin c(x)=\frac{\sin(\pi x)}{\pi x}$.
From Eq. (2), by the intensity correlation measurements, the best resolution of far-field GI with thermal light is determined by the speckle's
transverse size on the object plane ($\Delta x_s\approx \frac{{\lambda z}}{D}$), which is the same as GI in the
near field or Fresnel region \cite{Zhang,Ferri}.

For far-field GI scheme shown in Fig. 1(a), a single pointlike
detector far from the object is enough to record the complete
information of the object and its image in real-space can be
reconstructed by measuring the intensity correlation function
between the two detectors \cite{Gong}. According to Klyshko's
``advance optics'' picture \cite{Klyshko}, as shown in Fig. 1(b),
the object can be considered as being illuminated by a light source
emitting the light from the test detector $D_t$. After inverse
propagating in free space, the Fourier-transform (FT) diffraction
pattern of the object will appear in the far field of it (namely the
source plane $S$). Because the thermal source $S$ acts as a phase
conjugated mirror \cite{Valencia} and a spatial low-pass filter when
the source's transverse size is finite, only the low space frequency
part of the diffraction pattern will be reflected into the reference
path. After propagating along the reference path to the far field of
the source $S$, the reflected diffraction pattern will be inverse
Fourier-transformed and a low resolution real-space image will
finally be recorded by the camera $D_r$. However, for the case shown
in Fig. 1(b), Ref. \cite{Donoho} has proved theoretically that
super-resolution imaging can be obtained by exploiting the images'
sparsity constraints, and CS also utilizes the sparsity constraints
of the images in the recovery process while the image extraction
process of GI satisfies RIP of CS. Thus, combining GI with CS,
superresolved images can be reconstructed by ghost imaging via
sparsity constraints (GISC) described next in detail.

To GISC, we formulate it in the CS framework. For the GI system
shown in Fig. 1(a), each of the speckle intensity distributions on
the detection plane $D_r$ at time $s$ is described by $I_s(x,y)$
($n\times m$ pixels) and is reshaped as a row vector ($1\times N$,
$N=n\times m$). After $K$ measurements, the random sensing matrix
${\rm{\textbf{A}}}$ ($K\times N$) is reconstructed and at the same
time, the intensities ($B_s$) recorded by the test detector $D_t$
are arranged as a column vector ${\rm{\textbf{Y}}}$ ($K\times 1$).
If we denote the unknown object image as a $N$-dimensional column
vector ${\rm{\textbf{X}}}$ ($N\times 1$) and ${\rm{\textbf{X}}}$ can
be represented as ${\rm{\textbf{X}}}={\rm{\Psi}}{\rm{ \
}}{\rm{\cdot}}{\rm{\ }}{\rm{\alpha}}$ such that ${\rm{\alpha}}$ is
sparse in the representation basis ${\rm{\Psi}}$, then the image can
be reconstructed by solving the following convex optimization
program \cite{Candes,Figueiredo}:
\begin{eqnarray}
{\rm{X}}={\Psi\cdot\rm{\alpha}}; {\rm{ \ }} {\rm{ which \ minimizes:
}}{\rm{ \ }} \frac{{\rm{1}}}{{\rm{2}}}\left\|
{{{\rm{\textbf{Y}}}-\rm{\textbf{AX}}}} \right\|_2^2 + \tau \left\|
{\rm{\alpha}} \right\|_1.
\end{eqnarray}
where $\tau$ is a nonnegative parameter, $\left\| V \right\|_{2}$
denotes the Euclidean norm of $V$, and $\left\| V \right\|_{ 1 }  =
\sum\nolimits_i {\left| {\upsilon_i } \right|}$ is the $\ell_1$ norm
of $V$. Therefore, for the image with sparse cartesian
representation, the reconstruction process can be clearly written as
follows based on Eq. (3):
\begin{eqnarray}
\left| {T_{{\rm{GISC} }}} \right| = \left| {T'} \right|;{\rm{ \ }}
{\rm{ which \ minimizes: }}{\rm{ \ }}
\frac{{\rm{1}}}{{\rm{2}}}\left\| {B_s  - \int {dx } {dy } }
\right.\nonumber\\\times \left. {I_s (x,y)\left| {T'(x,y)} \right|^2
} \right\|_2^2  + \tau \left\| {\left| {T'(x,y)} \right|^2 }
\right\|_1 ,{\rm{ }}\forall _s  = 1 \cdots K.
\end{eqnarray}
where
\begin{eqnarray}
I_s (x,y) &\propto& \int {dx_1 } {dy_1 } {dx_2 } {dy_2 } E_s(x_1 ,y_1 )E_s^ *  (x_2 ,y_2 ) \nonumber\\&\times&\exp \{  -
\frac{{2j\pi }}{{\lambda z}}[(x_1  - x_2 )x + (y_1  - y_2 )y]\}.
\end{eqnarray}
\begin{eqnarray}
B_s &\propto& \int {dx_1 } {dy_1 } {dx_2 } {dy_2 }{dx} ' {dy'} {dx} '' {dy} ''T(x',y') T^ *(x'',y'')\nonumber\\&\times& E_s(x_1 ,y_1 )
\exp \{ \frac{{j\pi }}{{\lambda z}}({x'}^2+ {y'}^2-{x''}^2-{y''}^2)\}\nonumber\\&\times& E_s^ *  (x_2 ,y_2 )\exp \{ \frac{{j\pi }}{{\lambda z_1 }}
(x'^2 + y'^2 - x''^2- y''^2 )\}\nonumber\\&\times& \exp \{ \frac{{2j\pi }}{{\lambda z}}(x'' x_2+y'' y_2-x'x_1-y'y_1)\}\nonumber\\&\times&
\sin c[\frac{{L_1 }}{{\lambda z_1 }}(x' - x'')]\sin c[\frac{{L_1 }}{{\lambda z_1 }}(y' - y'')].
\end{eqnarray}
and $\left| {T_{{\rm{GISC} }}} \right|$ is the object's transmission
function recovered by GISC method, $L_1$ is the effective receiving
aperture of the test detector $D_t$.

Figs. 2-3 present experimental results of a double-slit recovered by
GI and GISC methods in different receiving areas $L_1\times L_1$ and
different distances $z_1$, using the schematic shown in Fig. 1(a).
For GISC method, we have utilized the gradient projection for sparse
reconstruction algorithm \cite{Figueiredo,Website}. As shown in Fig.
2(c-d), the object's image can not be reconstructed by GI because
the transverse size of the speckle on the object plane $\Delta x_s$
is much larger than center-to-center separation of the object, which
is consistent with the result expressed by Eq. (2) and accords with
the physical explanation described in Fig. 1(b). However, the images
with the resolution beyond $\frac{1}{6}\Delta x_s$ (Fig. 2(e,f)) and
Fig. 3(a-d)) can be obtained by GISC. As the receiving areas of the
detector $D_t$ are increased or the distance between the object and
the detector $D_t$ is decreased, the quality of GISC will be
improved (Fig. 2(e,f) and Fig. 3(a-d)), which can be explained by
Eqs. (4-6) because the Euclidean term in Eq. (4) will approach zero
such that Eq. (4) becomes the linear $\ell_1$-norm problem as the
increase of $L_1$ or the decrease of $z_1$ \cite{Candes}. Moreover,
as shown in Fig. 2(e,f), the resolution of the images reconstructed
by GISC also depends on the pixel-resolution of the camera $D_r$.

\begin{figure}
\centerline{
\includegraphics[width=8.5cm]{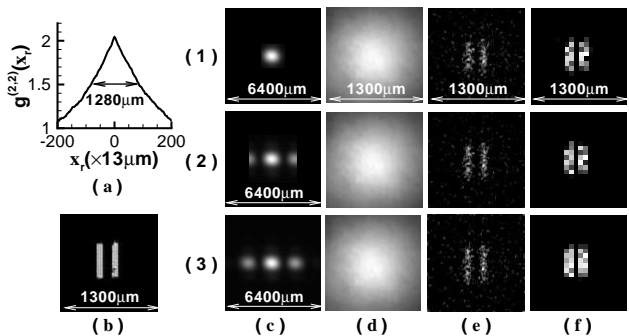}}
\caption{Experimental reconstruction of a double-slit in different
receiving areas with $z$=1200mm, $z_1$=500mm and $D$=0.6mm (the
speckle's transverse size on the object plane $\Delta x_s$=1280mm).
(a). The cross-section curve of the speckle on the object plane
obtained by measuring the second-order correlation function of light
field on the reference detection plane (the curve's full-width at
half-max is the resolution limitation of GI); (b). the object; (c).
the object's diffraction patterns received by the test detector
$D_t$; (d). GI method (averaged 3000 measurements); (e) and (f) are
GISC when the pixel-resolution of the camera $D_r$ is 13$\mu$m and
65$\mu$m, respectively (with 3000 and 500 measurements for (e)-(f),
respectively). The receiving areas of the detector $D_t$ shown in
(1-3) are 1.6mm$\times$1.6mm, 3.2mm$\times$3.2mm, and
6.4mm$\times$6.4mm.}
\end{figure}

\begin{figure}
\centerline{
\includegraphics[width=8.5cm]{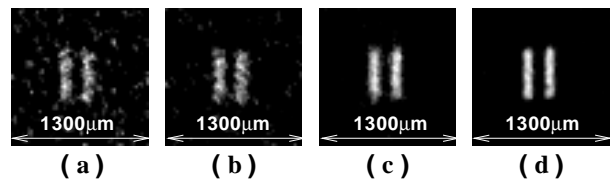}}
\caption{Experimental results of the same double-slit in different
distances $z_1$, and the other conditions are the same as Fig. 2
(using 1000 measurements). (a)-(d) are GISC when the
pixel-resolution of the camera $D_r$ is 26$\mu$m and $z_1$=500mm,
200mm, 100mm and 10mm, respectively. The receiving area of the
detector $D_t$ is 6.4mm$\times$6.4mm.}
\end{figure}

To verify the super-resolution ability of GISC for more general
images and the effect of the object's sparse representation on the
quality of GISC, as shown in Fig. 4(c) and (e), a transmission
aperture (``\textbf{zhong}''ring) with the resolution beyond
$\frac{1}{5}\Delta x_s$ is reconstructed by GISC in cartesian and
DCT representation basis, which suggests that the images with much
better quality can be obtained by choosing a proper representation
basis. Therefore, for the first time, we demonstrate experimentally
that far-field superresolved imaging can be realized by utilizing
the images' sparsity constraints in ghost imaging schemes.

\begin{figure}
\centerline{
\includegraphics[width=8.5cm]{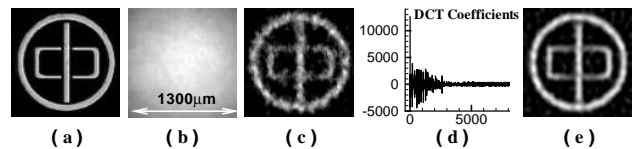}}
\caption{Recovered results of an aperture (``\textbf{zhong}''ring)
in different representation basis, with the same conditions of Fig.
3 and $z_1$=10mm (using 2000 measurements). (a). The object; (b). GI
reconstruction; (c). GISC reconstruction when the object is
represented in cartesian basis; (d). the object's discrete cosine
transform (DCT) coefficients; and (e). GISC reconstruction when the
object is represented in DCT basis. }
\end{figure}

In single-photon imaging system, each of the photons only interferes
with itself \cite{Dirac}, it is impossible to obtain the real-space
image of a double-slit and its diffraction pattern at the same time
because a photon cannot pass both of the slits to generate the
double-slit's diffraction pattern while at the same time pass one of
them to give out the double-slit's image in real-space. For GI,
based on the property of spatial correlation between two light
fields, it is also impossible to obtain both the image in real-space
of the double-slit and its diffraction pattern at the same time in
fixed GI schemes \cite{Ferri,Bache,Basano}. However, by taking the
image's sparsity as a priori information, in far-field GISC system
shown in Fig. 1(a), when the transverse size of the speckle on the
object plane is much larger than center-to-center separation of the
double-slit and the test detection plane is located in the far field
of the double-slit, the double-slit's diffraction pattern and its
real-space image, as shown in Fig. 2(c) and Fig. 2(e,f), can be
obtained at the same time. Moreover, the reconstruction results of
GISC don't only depend on how we measure the object as in a standard
quantum measurement frames, but also depend on how sparse the object
is in the representation basis (Fig. 3(d) and Fig. 4(c,e)).
Actually, for any GI system, we can find a suitable representation
basis in which the object is sufficiently sparse, therefore, as
shown in Fig. 3(d) and Fig. 4(c,e), super-resolution imaging can be
achieved and GISC will be a universal super-resolution imaging
method. Understanding what happens at quantum level in GISC seems to
be an interesting challenge deserving more investigation.

In conclusion, we have achieved super-resolution far-field GI by
combining GI method with the sparsity constraints of images. Both
the approaches to realize the linear $\ell_1$-norm problem and an
optimal representation basis can dramatically enhance the image's
reconstruction quality. We have also shown that Fourier-transform
diffraction pattern of the object and its image in real-space can be
obtained at the same time. This brand new far-field super-resolution
imaging method will be very useful to microscopy in biology,
material, medical sciences, and in the filed of remote sensing, etc.

The work was supported by the Hi-Tech Research and Development
Program of China under Grant Project No. 2011AA120101 and No.
2011AA120202.

\end{document}